\documentclass[runningheads]{svmult}

\usepackage{makeidx}   
\usepackage{graphicx}  
\usepackage{subeqnar}  
\usepackage{multicol}  
\usepackage{physprbb}  
\makeindex             

\newcommand{\greeksym}[1]{{\usefont{U}{psy}{m}{n}#1}}
\newcommand{\umu}{\mbox{\greeksym{m}}}

\begin{document}
\title*{The Innermost AGNs with Future mm-VLBI}
\toctitle{The Innermost AGNs with Future mm-VLBI}
%
%
\titlerunning{The Innermost AGNs with Future mm-VLBI}
%

\author{I.~Agudo\inst{1}
\and T.P.~Krichbaum\inst{1}
\and U.~Bach\inst{1}
\and A.~Pagels\inst{1}
\and B.W.~Sohn\inst{1}
\and D.A.~Graham\inst{1}
\and A.~Witzel\inst{1}
\and J.A.~Zensus\inst{1}
\and J.L.~G\'omez\inst{2,3}
\and M.~Bremer\inst{4}
\and M.~Grewing\inst{4}}
\authorrunning{I. Agudo et al.}
%
%
\institute{MPIfR, Auf dem H\"ugel 69, 53121 Bonn, Germany
\and IAA(CSIC), Apartado 3004, 18080 Granada, Spain
\and IEEC/CSIC, Edifici Nexus, C/Gran Capit\`a, 2-4, E-08034 Barcelona, Spain
\and IRAM, Grenoble, 300 Rue de la Piscine, 38406 Saint Martin d'H\`eres, France}

\maketitle              

\section{mm-VLBI, Astronomy at the Highest Resolution}
More than 40 years since the discovery of the AGNs, there are
still fundamental questions related to the nature of these
intriguing objects. In particular, the accretion processes
onto their super-massive black holes and the mechanisms
through which their relativistic jets are formed, accelerated
and collimated are still not well understood. Great effort
has been made during the last decade to push the mm-VLBI
technique to progressively shorter wavelengths, offering
the best tool to observe the innermost regions of the
jets and study the physics involved in their behaviour.

At present, the most sensitive mm-VLBI instrument is the
Global mm-VLBI Array, composed of the Effelsberg, Plateau
de Bure, Pico Veleta, Onsala and Mets\"ahovi stations, in
addition to eight of the ten VLBA antennas (for more details
see \emph{http://www.mpifr-bonn.mpg.de/div/vlbi/globalmm}).
The Global mm-VLBI Array reaches a baseline
sensitivity of 80--100\,mJy (adopting 20\,s coherence time,
100\,s segmentation time and a sampling rate of 512\,Mbps
[2bits]). This yields an image sensitivity of $1$--$2$\,mJy
(for 12\,h of observation and a duty cycle of 0.5). With these
characteristics the number of sources which could be imaged
with high dynamic ranges ($\ge$100:1) is nowadays larger
than 100.

\begin{figure}[b]
\begin{center}
\includegraphics[width=0.51\textwidth]{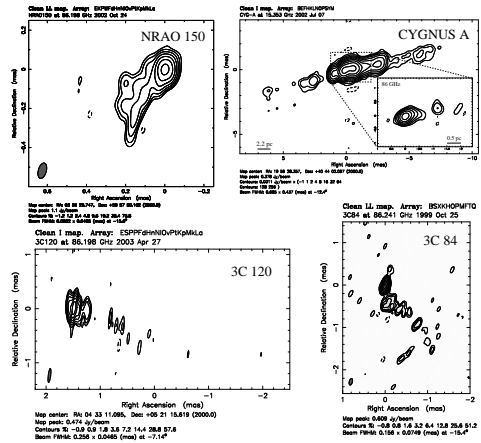}
\end{center}
\caption[]{3mm-VLBI images of NRAO150, Cygnus A, 3C~120 and 3C~84.}
\label{F1}
\end{figure}

In an attempt to obtain a deeper knowledge of the physics
in the innermost regions of jets in AGNs, we have started
a VLBI monitoring, at 3mm, of some of the brighter-most
sources. Fig.~\ref{F1} represents some of the images from
these observations (still in progress). The images
demonstrate the capability of the Global mm-VLBI Array
to study the innermost jet structures with an \emph{angular
resolution better than 50\,$\umu$as}.

\section{The Future: Higher Sensitivity and Image Fidelity}
In order to achieve a better quality of images and increase the
number of sources that can be observed, a further increase in
sensitivity is still needed. To do that, the most direct way
is to increase the collecting area of the present interferometer.
For the near future, ALMA, the GBT, the LMT, CARMA, SRT, Yebes,
Nobeyama and Noto are some of the most sensitive stations
suitable to participate in mm-VLBI. This future array, together
with the present Global mm-VLBI Array, would achieve baseline
sensitivities of up to 5--10\,mJy (assuming 1\,Gbps recording rate
and 100\,s segmented integration time), and an image sensitivity
better than 0.1\,mJy. These estimates predict a large
\emph{increase, by a factor of 10!}, respect to the present
Global mm-VLBI Array levels of sensitivity. In addition,
continuous development of VLBI will provide standard recording
rates of at least 2\,Gbps in the next years, which will increase
the expected sensitivities by an extra factor $\ge\sqrt{2}$.
Further significant improvements in coherence time can be
reached by atmospheric phase correction methods.

But the proposed future array will not only influence the
sensitivity. The new stations will also largely improve the
UV-coverage, and so the image fidelity.
The addition of ALMA will improve the UV-coverage for
sources with low declination (less than $20^{\circ}$) and
facilitate the VLBI imaging of the Galactic Centre source
SgrA*. With these improvements, dynamic ranges
of $\ge$1000:1 could be easily obtained. This will
place mm-VLBI at comparable levels of sensitivity and image
fidelity than present day cm-VLBI.

\section{Science with Future mm-VLBI}
The expected improvements in sensitivity and image fidelity
would impact our knowledge of the physics of jets and central
engines in AGNs. It would be possible to obtain high
quality images of the innermost regions in the jets. This
would facilitate, for several hundreds or even thousands of
compact sources (i) to investigate the MHD physics in
strong gravitational fields, (ii) to study the formation,
initial acceleration and collimation of relativistic
jets, (iii) to probe their initial magnetic field
configurations (via polarimetry) and (iv) to infer the
properties of the super-massive black holes and their
immediate vicinity.

%

\end{document}